\documentclass[12pt]{article}
\usepackage{epsfig}
\usepackage{amssymb}
\usepackage{amsmath}
\usepackage{amsfonts}
\usepackage{graphicx}
\usepackage{mathrsfs}
\DeclareMathAlphabet{\mathscrbf}{OMS}{mdugm}{b}{n}
\usepackage{mathabx}
\usepackage[dvips]{color}
\usepackage{multirow}
\usepackage{calc}
\usepackage{accents}

\newcommand{\bGamma}{{\boldsymbol{\Gamma}}}
\newcommand{\bsigma}{{\boldsymbol{\sigma}}}

\newcommand{\bnabla}{\boldsymbol{\nabla}}

\newcommand{\R}{\mathbb{R}}
\newcommand{\C}{\mathbb{C}}
\newcommand{\Z}{\mathbb{Z}}

\newcommand{\ff}{\mathfrak{f}}
\newcommand{\fg}{\mathfrak{g}}

\newcommand{\fz}{\mathfrak{z}}

\newcommand{\fF}{\mathfrak{F}}

\newcommand{\fK}{\mathfrak{K}}

\newcommand{\bfe}{\mathbf{e}}
\newcommand{\bff}{\mathbf{f}}

\newcommand{\bh}{\mathbf{h}}
\newcommand{\bk}{\mathbf{k}}

\newcommand{\bfr}{\mathbf{r}}

\newcommand{\bv}{\mathbf{v}}

\newcommand{\bE}{\mathbf{E}}
\newcommand{\bcE}{{\boldsymbol{\cE}}}
\newcommand{\bF}{\mathbf{F}}

\newcommand{\bH}{\mathbf{H}}
\newcommand{\bI}{\mathbf{I}}

\newcommand{\bK}{\mathbf{K}}
\newcommand{\bL}{\mathbf{L}}

\newcommand{\bM}{\mathbf{M}}

\newcommand{\bT}{\mathbf{T}}
\newcommand{\bU}{\mathbf{U}}
\newcommand{\bV}{\mathbf{V}}

\newcommand{\cH}{\mathcal{H}}
\newcommand{\cE}{\mathcal{E}}
\newcommand{\cF}{\mathcal{F}}

\newcommand{\cS}{\mathcal{S}}

\newcommand{\be}{\begin{equation}}
\newcommand{\ee}{\end{equation}}
\newcommand{\bea}{\begin{eqnarray}}
\newcommand{\eea}{\end{eqnarray}}
\newcommand{\nn}{\nonumber}
\newcommand{\kt}{\rangle}
\newcommand{\br}{\langle}

\newcommand{\ed}{\end{document}}

\newcommand{\bi}{\begin{itemize}}
\newcommand{\ei}{\end{itemize}}

\newcommand{\bce}{\begin{center}}
\newcommand{\ece}{\end{center}}

\newcommand{\sD}{\mathscr{D}}

\newcommand{\sF}{\mathscr{F}}

\newcommand{\bsH}{\mathscrbf{H}}

\newcommand{\sV}{\mathscr{V}}

\newcommand{\bsU}{\mathscrbf{U}}
\newcommand{\RE}{{\rm Re}}

\newcommand{\bPhi}{{\boldsymbol{\Phi}}}
\newcommand{\bPi}{{\boldsymbol{\Pi}}}

\newcommand{\bcH}{{\boldsymbol{\cH}}}

\newcommand{\bvarepsilon}{{\mbox{\large$\boldsymbol{\varepsilon}$}}}
\newcommand{\bmu}{{\mbox{\large$\boldsymbol{\mu}$}}}

\newcommand{\bigvarepsilon}{\mbox{\large$\varepsilon$}}
\newcommand{\bigmu}{\mbox{\large$\mu$}}

\newcommand{\bzero}{{\boldsymbol{0}}}

\newcommand{\for}{{\mbox{\rm for}}}

\newcommand{\bup}{{\boldsymbol{\Upsilon}}}
\newcommand{\bXi}{{\boldsymbol{\Xi}}}
\newcommand{\bfeta}{{\boldsymbol{\eta}}}

\newcommand{\Lpi}{{\widehat{\boldsymbol{\mbox{\Large${{\pi}}$}}}_{\!k}}}
\newcommand{\sLpi}{{\widehat{{\mbox{\Large${{\pi}}$}}}_{\!k}}}

\newcommand{\bfvarepsilon}{{\boldsymbol{\varepsilon}}}
\newcommand{\bfmu}{{\boldsymbol{\mu}}}

\newcommand{\bsfE}{{\boldsymbol{\mathsf{E}}}}
\newcommand{\bsfH}{{\boldsymbol{\mathsf{H}}}}

\begin{document}

\title{Exactness of the first Born approximation in electromagnetic scattering}


\author{Farhang Loran\thanks{E-mail address: loran@iut.ac.ir}
~and Ali~Mostafazadeh\thanks{Corresponding author, E-mail address:
amostafazadeh@ku.edu.tr}\\[6pt]
$^*$Department of Physics, Isfahan University of Technology, \\ Isfahan 84156-83111, Iran\\[6pt]
$^\dagger$Departments of Mathematics and Physics, Ko\c{c} University,\\  34450 Sar{\i}yer,
Istanbul, T\"urkiye
}

\date{ }
\maketitle

\begin{abstract}
For the scattering of plane electromagnetic waves by a general possibly anisotropic stationary linear medium in three dimensions, we give a condition on the permittivity and permeability tensors of the medium under which the first Born approximation yields the exact expression for the scattered wave whenever the incident wavenumber $k$ does not exceed a pre-assigned value $\alpha$. We also show that under this condition the medium is omnidirectionally invisible for $k\leq \alpha/2$, i.e., it displays broadband invisibility regardless of the polarization of the incident wave.

\vspace{2mm}



\end{abstract}

\section{Introduction}
\label{S1}

Since its inception in 1926 \cite{Born-1926}, the Born approximation \cite{born-wolf,taylor,newton}  has been the principal approximation sheme for performing scattering calculations  \cite{hofstadter,Breuer-1981,TKD,Abubakar-2005,koshino,Hunter-2006, Bennett-2014,Bereza-2017,van der Sijs}. Yet the search for scattering systems for which the Born approximation is exact did not succeed until 2019 where the first examples of complex potentials possessing this property could be constructed within the context of potential scattering of scalar waves in two dimensions \cite{pra-2019}. The key ingredient leading to the discovery of these potentials is a recently proposed dynamical formulation of stationary scattering \cite{pra-2016,prsa-2016,pra-2021}. The purpose of the present article is to employ the dynamical formulation of electromagnetic scattering developed in \cite{pra-2023} to address the problem of the exactness of the first Born approximation in electromagnetic scattering.

Consider the scattering of plane electromagnetic waves by a general stationary linear medium. The electric field $\boldsymbol{\mathsf{E}}$ of the wave, which together with its magnetic field  $\boldsymbol{\mathsf{H}}$ satisfy Maxwell's equations, admits the asymptotic expression: $\boldsymbol{\mathsf{E}}(\bfr,t)=\boldsymbol{\mathsf{E}}_{\rm i}(\bfr,t)+
	\boldsymbol{\mathsf{E}}_{\rm s}(\bfr,t)$ for $r\to\infty$,
where $\boldsymbol{\mathsf{E}}_{\rm i}$ and $\boldsymbol{\mathsf{E}}_{\rm s}$ are respectively the electric fields of the incident and scattered waves, $\bfr$ is the position of the detector observing the wave, and $r:=|\bfr|$. These fields have the form \cite{newton,TKD}:
	\bea
	\boldsymbol{\mathsf{E}}_{\rm i}(\bfr,t)&=&
	{\mathsf{E}}_0\;e^{i(\bk_{\rm i}\cdot\bfr-\omega t)}\:\bfe_{\rm i},\\
	\boldsymbol{\mathsf{E}}_{\rm s}(\bfr,t)&=&
	\frac{{\mathsf{E}}_0\;e^{i(kr-i\omega t)}}{r}\: 
	\bF(\bk_{\rm s},\bk_{\rm i}),
	\label{Scattered-wave-lr}
	\eea
where ${\mathsf{E}}_0$ is a complex amplitude, and $\bk_{\rm i}$, $\omega$, and $\bfe_{\rm i}$ are respectively the wave vector, angular frequency, and polarization vector of the incident wave, $k:=\omega/c=|\bk_{\rm i}|$ is the wavenumber, $\bF$ is a vector-valued function, $\bk_{\rm s}:=k\hat\bfr$, and $\hat\bfr:=r^{-1}\bfr$.

The electric field of the scattered wave turns out to admit a perturbative series expansion known as the Born series \cite{born-wolf,newton,TKD}. We can express it as 
	\be
	\boldsymbol{\mathsf{E}}_{\rm s}(\bfr,t)=
	\frac{{\mathsf{E}}_0\;e^{i(kr-i\omega t)}}{r}
	\sum_{n=1}^\infty \bF_n(\bk_{\rm s},\bk_i),
	\label{Born-series}
	\ee
where $\bF_n$ are vector-valued functions \cite{Kilgore-2017}. The $N$-th order Born approximation amounts to neglecting all but the first $N$ terms of the series in (\ref{Born-series}).

To reveal the perturbative nature of the Born series, we introduce:
$\bfeta_\bfvarepsilon(\bfr):=\hat{\bfvarepsilon}(\bfr)-\bI$ and $\bfeta_\bfmu(\bfr):=\hat{\bfmu}(\bfr)-\bI$, where $\hat\bfvarepsilon$ and $\hat\bfmu$ are respectively the relative permittivity and permeability tensors\footnote{Recall that ${\hat\bfvarepsilon}:=\varepsilon_0^{-1}\bfvarepsilon$ and $\hat\bfmu\,:=\mu_0^{-1}\bfmu$, $\bfvarepsilon$ and $\bfmu$ denote the  permittivity and permeability tensors, and $\varepsilon_0$ and $\mu_0$ are the permittivity and permeability of the vacuum.} of the medium, and $\bI$ is the $3\times 3$ identity matrix. Let $\varsigma$ be a positive real number. Then under the scaling transformation,
	\begin{align}
	&\bfeta_\bfvarepsilon(\bfr)\to\varsigma\,\bfeta_\bfvarepsilon(\bfr),
	&&\bfeta_\bfmu(\bfr)\to\varsigma\,\bfeta_\bfmu(\bfr),
	\label{scaling}
	\end{align}
the vector-valued functions $\bF_n$, which determine the terms of the Born series (\ref{Born-series}), transform as\footnote{This follows from the recurrence relation for $\bF_n$ which has its root in the structure of the electromagnetic Lippmann-Schwinger equation \cite{Kilgore-2017}.}
	\be
	\bF_n(\bk_{\rm s},\bk_{\rm i})\to\varsigma^n\bF_n(\bk_{\rm s},\bk_{\rm i}).
	\label{scaling-trans}
	\ee
Because $\bfeta_\bfvarepsilon(\bfr)$ and $\bfeta_\bfmu(\bfr)$ quantify the scattering properties of the medium, the transformations (\ref{scaling}) with $\varsigma<1$ correspond to a medium with a weaker scattering response. This in turn shows that for such a medium, $|\bF_n(\bk_{\rm s},\bk_{\rm i})|$ becomes increasing small as $n$ grows, and terminating the Born series yields a reliable approximation. The principal example is the first Born approximation which involves neglecting all but the first term of the Born series \cite{born-wolf,newton,TKD}. This approximation is exact if 
	\be
	\bF_n(\bk_{\rm s},\bk_{\rm i})=\bzero~~~~\for~~~~n\geq 2,
	\label{exact-2}
	\ee
equivalently
	\be
	\boldsymbol{\mathsf{E}}_{\rm s}(\bfr,t)=
	\frac{{\mathsf{E}}_0\;e^{i(kr-i\omega t)}}{r}\:
	\bF_1(\bk_{\rm s},\bk_{\rm i}).
	\label{exact}
	\ee
We can also express this condition in terms of the scaling rule (\ref{scaling-trans}); we state it as a theorem for later reference.
	\begin{itemize}
	\item[]{\bf Theorem~1}: The first Born approximation is exact if and only if under the scaling transformation (\ref{scaling}), the electric field of the scattered wave transforms as $\boldsymbol{\mathsf{E}}_{\rm s}\to\varsigma\:\boldsymbol{\mathsf{E}}_{\rm s}$.
	\end{itemize}
	
	Given the difficulties associated with finding explicit formulas for $\bF_n(\bk_{\rm s},\bk_{\rm i})$ and the fact that (\ref{exact-2}) corresponds to an infinite system of complicated integral equations (constraints) for $\hat\bfvarepsilon$ and $\hat\bfmu$, it is practically impossible to use (\ref{exact-2}) for the purpose of determining the  permittivity and permeability profiles for which the first Born approximation is exact.\footnote{The transverse vector nature of the electromagnetic waves and the tensorial nature of the corresponding interaction potentials $\bfeta_\bfvarepsilon$ and $\bfeta_\bfmu$ make this a considerably more elaborate task than addressing the same problem for scalar waves.} This is the main reason why identifying the explicit conditions for the exactness of the first Born approximation has been an open problem for close to a century.\footnote{There have been extensive studies of the Born series and its convergence properties in quantum scattering theory of scalar waves \cite{jost-1951,Kohn-1954,Zemach-1958,Aaron-1960,corbett-1968,Bushell-1972} as well as the scattering theory of electromagnetic waves \cite{Kilgore-2017}. None of these, however, provide conditions for the truncation of this series.} Motivated by our results on the scattering of scalar waves \cite{pra-2019}, we pursue a different route toward a solution of this problem. This is based on a dynamical formulation of the stationary electromagnetic scattering \cite{pra-2023} whose main ingredient is a fundamental notion of transfer matrix. This is a linear operator acting in an infinite-dimensional function space that similarly to the traditional numerical transfer matrices \cite{teitler-1970,berreman-1972,pendry-1984,pendry-1990a,mclean,ward-1996,pendry-1996} stores the information about the scattering properties of the medium but unlike the latter allows for analytic calculations. In this article, we use the fundamental transfer matrix to obtain a sufficient condition for the exactness of the first Born approximation in electromagnetic scattering. 
		
The outline of this article is as follows. In Sec.~2 we present our main results as well as specific examples of scattering media for which the first Born approximation is exact. In Sec.~3 we offer a concise review of dynamical formulation of the stationary electromagnetic scattering. In Sec.~4, we discuss the application of this formulation in addressing the problem of finding conditions for the exactness of the first Born approximation. In Sec.~5, we present a summary of our findings and our concluding remarks. 

\section{Main results}
	
\begin{figure}
        \begin{center}
        \includegraphics[scale=.35]{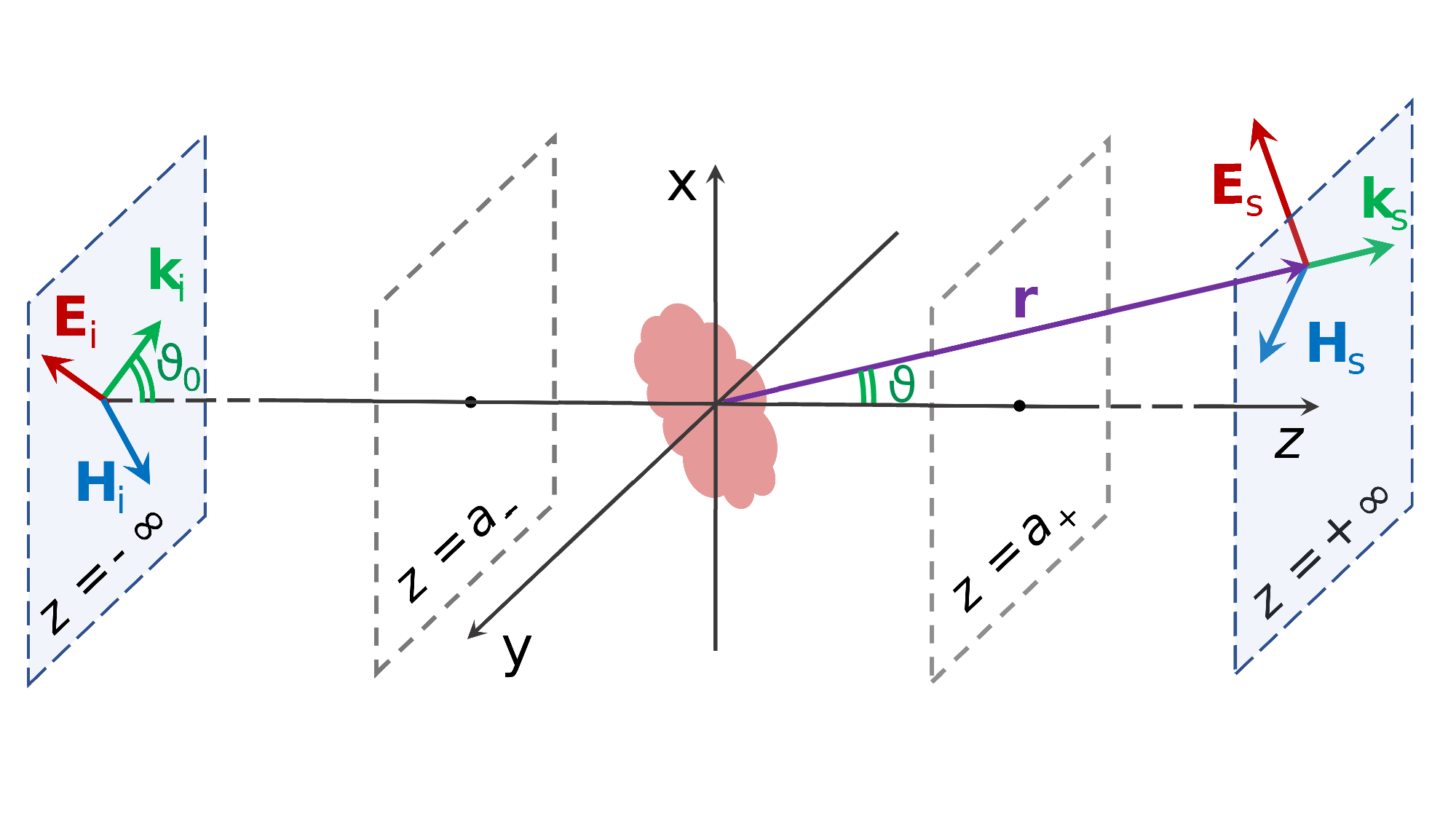}\vspace{-1cm}
        \caption{Schematic view of scattering setup where the source of the incident wave lies on the plane $z=-\infty$. The region colored in pink represents the scatterer that is confined between the planes $z=a_\pm$. $\bk_{\rm i}$, $\bE_{\rm i}$, and $\bH_{\rm i}$ are respectively the wave vector, the electric field, and the magnetic field of the incident wave, while $\bk_{\rm s}$, $\bE_{\rm s}$, and $\bH_{\rm s}$ are respectively the wave vector, the electric field, and the magnetic field of the scattered wave detected by a detector placed on the plane $z=+\infty$.}
        \label{fig1}
        \end{center}
        \end{figure}

We begin our analysis by considering the scattering setup where the source of the incident wave and the detectors detecting the scattered wave are, without loss of generality, placed on the planes $z=\pm\infty$ in a Cartesian coordinate system with coordinates $x,y$, and $z$, as depicted in Fig.~\ref{fig1}. 
We also suppose that the space outside the region bounded by a pair of normal planes to the $z$ axis is empty, i.e., there is an interval $(a_-,a_+)$ on the $z$ axis such that  
	\be
	\bfeta_\bfvarepsilon(x,y,z)=
	\bfeta_\bfmu(x,y,z)=\bzero~~~\for~~~z\notin (a_-,a_+).
	\label{condi-0}
	\ee
Furthermore, we assume that the Fourier transform of all functions of real variables that enter our analysis exists. 
	
Throughout this article we employ the following notations.
	\begin{itemize}
	\item[-] For each vector $\bv\in\R^3$, $v_x,v_y$, and $v_z$ label the $x$, $y$, and $z$ components of $\bv$, and $\vec v$ stands for $(v_x,v_y)$ so that $\bv=(\vec v,v_z)$. In particular, $\vec r:=(x,y)$ and $\bfr=(\vec r,z)$.
	\item[-] Given a scalar, vector-valued, or matrix-valued function $g$ of $\bfr$, we use $\tilde g(\vec p,z)$ to denote the two-dimensional Fourier transform of $g(\vec r,z)$ with respect to $\vec r$, i.e.,
	\be
	\tilde g(\vec p,z):=\int_{\R^2}d^2 r\: e^{-i\vec p\cdot\vec r}g(\vec r,z)=\int_{-\infty}^\infty dx\int_{-\infty}^\infty dy\: e^{-i(xp_x+yp_y)}g(x,y,z),
	\label{Fourier}
	\ee
where $\vec p:=(p_x,p_y)$. For example, $\tilde{\bfeta}_\varepsilon$ and $\tilde{\bfeta}_\mu$ are respectively the two-dimensional Fourier transforms of ${\bfeta}_\varepsilon$ and ${\bfeta}_\mu$ with respect to $\vec r$.
	\item[-] We use $\hat\varepsilon_{ij}$ and $\hat\mu_{ij}$ to mark the entries of $\hat\bfvarepsilon$ and $\hat\bfmu$, respectively.	
	\end{itemize}
The following is the main result of this article which we prove in Sec.~4.
	\begin{itemize}
	\item[]{\bf Theorem~2} Consider the electromagnetic scattering problem for a time-harmonic plane wave propagating in a stationary linear medium with
relative permittivity and permeability, $\hat\bfvarepsilon$ and $\hat\bfmu$. Suppose that ${\bfeta}_\varepsilon:=\hat\bfvarepsilon-\bI$ and ${\bfeta}_\mu:=\hat\bfmu-\bI$ satisfy (\ref{condi-0}) for some $a_\pm\in\R$ with $a_-<a_+$ and that the following conditions hold.
 	\begin{enumerate}
	\item $\hat\varepsilon_{33}$ and $\hat\mu_{33}$ are bounded functions whose real part has a positive lower bound, i.e., there are real numbers $m$ and $M$ such that for all $\bfr\in\R^3$, 
		\begin{align}
		&0<m\leq\RE[\hat\varepsilon_{33}(\bfr)]\leq |\hat\varepsilon_{33}(\bfr)|\leq M,
		&&0<m\leq\RE[\hat\mu_{33}(\bfr)]\leq |\hat\mu_{33}(\bfr)|\leq M,
		\label{bound}
		\end{align}	
where ``$\RE$'' stands for the real part of its argument.
	\item There are a positive real number $\alpha$ and a unit vector $\vec e$ lying on the $x$-$y$ plane such that 
		\be
		\tilde{\bfeta}_\bfvarepsilon(\vec p,z)=		
		\tilde{\bfeta}_\bfmu(\vec p,z)=\bzero~~~~\for~~~~
		\vec p\cdot\vec e\leq \alpha.
		\label{condi}
		\ee
	\end{enumerate}
Then the first Born approximation provides the exact solution of the scattering problem, if the wavenumber $k$ for the incident wave does not exceed $\alpha$, i.e., $k\leq \alpha$. Moreover, the medium does not scatter the incident waves with wavenumber $k\leq \alpha/2$, i.e., it displays broadband invisibility in the wavenumber spectrum $(0,\alpha/2]$.
	\end{itemize}
Notice that Condition~1 of this theorem holds for all non-exotic isotropic media. We can also satisfy it for realistic anisotropic media by an appropriate choice of the $z$ axis. Furthermore, if Condition~2 holds, we can perform a rotation about the $z$ axis to align $\vec e$ and the $x$ axis in which case (\ref{condi}) takes the form
 	\be
	\tilde{\bfeta}_\bfvarepsilon(p_x,p_y,z)=		
	\tilde{\bfeta}_\bfmu(p_x,p_y,z)=\bzero~~~~\for~~~~
	p_x\leq \alpha.
	\label{condi2}
	\ee
Since such a rotation will not affect the two-dimensional Fourier transform of a function with respect to $x$ and $y$, Condition~2 is equivalent to (\ref{condi2}).

To provide concrete examples of linear media satisfying (\ref{condi2}), we confine our attention to nonmagnetic isotropic media, where $\hat\bfmu=\bI$, $\hat\bfvarepsilon=\hat\varepsilon(\bfr)\bI$, and $\hat\varepsilon$ is the scalar relative permittivity. Then $\bfeta_\bfmu=\bzero$ and $\bfeta_\bfvarepsilon=\eta_\varepsilon(\bfr)\bI$, where $\eta_\varepsilon=\hat\varepsilon-1$, and (\ref{condi2}) reduces to 
$\tilde{\eta}_\varepsilon(p_x,p_y,z)=0$ for $p_x\leq \alpha$. We can identify this with the condition that the Fourier transform of $e^{-i\alpha x}\eta_\varepsilon(x,y,z)$ with respect to $x$ vanishes on the negative $p_x$ axis.\footnote{In one dimension, scattering potentials with this property are known to be unidirectionally invisible for all wavenumbers \cite{horsley-2015,longhi-2015,horsley-longhi,jiang-2017}.} This means that there is a function $u:\R^3\to\C$ such that\footnote{To ensure the existence of the two-dimensional Fourier transform of $\hat\varepsilon-1$ with respect to $\vec r$, we can require that $\int_0^\infty d\fK\int_{-\infty}^\infty dy\:|u(\fK,y,z)|^2<\infty$ for all $z\in\R$.} 
	\be
	\hat\varepsilon(x,y,z)=1+ e^{i\alpha x} \int_0^\infty d\fK\; e^{i\fK x } u(\fK,y,z).
	\label{epsilon=}
	\ee
A class of possible choices for $u$, which allow for the analytic evaluation of the integral in (\ref{epsilon=}), is given by
	\be
	u(\fK,y,z)=\frac{a}{m!}\,(a\fK)^m e^{-a\fK}f(y,z), 
	\label{u=}
	\ee
where $a$ is a positive real parameter,  $m$ is a positive integer, and $f:\R^2\to\C$ is a function.\footnote{Requiring $\int_\infty^\infty dy|f(y,z)|^2<\infty$, we can guarantee the existence of the right-hand side of (\ref{epsilon=}) and the two-dimensional Fourier transform of $\hat\varepsilon-1$ with respect to $\vec r$.} Substituting (\ref{u=}) in (\ref{epsilon=}), we find
	\be
	\hat\varepsilon(x,y,z)=1+ \frac{e^{i\alpha x}f(y,z)}{(1-\frac{i x}{a})^{m+1}}.
	\label{epsilon=2}
	\ee
It is easy to show that (\ref{epsilon=2}) satisfies the first relation in (\ref{bound}), if there is a real number $b$ such that\footnote{Because $|1-ix/a|\geq 1$, (\ref{epsilon=2}) and (\ref{condi-001}) imply $|\eta_\varepsilon|=|\varepsilon-1|\leq|f|\leq b$. Using this relation  together with $-|\eta_\varepsilon|\leq\RE(\eta_\varepsilon)\leq|\eta_\varepsilon|$, $\RE(\hat\varepsilon)=1+\RE(\eta_\varepsilon)$, and $b<1$, we have $0<1-b\leq\RE(\hat\varepsilon)\leq 1+b$, i.e., (\ref{bound}) holds for $m=1-b$ and $M=1+b$.}
	\be
	|f(y,z)|\leq b<1~~~\mbox{for all}~~~(y,z)\in\R^2.
	\label{condi-001}
	\ee
Suppose that this condition holds. Then according to Theorem~2, the first Born approximation provides the exact solution of the scattering problem for the permittivity profile (\ref{epsilon=2}), if the incident wave has a wavenumber $k$ not greater than $\alpha$. Furthermore, the medium in invisible if $k\leq\alpha/2$. Another example for such a permittivity profile is
	\be
	\hat\varepsilon(x,y,z)=1+ \sqrt{\pi}\,e^{i\alpha x} e^{-\frac{x^2}{a^2}}\left[1+{\rm erf}\mbox{$(\frac{ix}{a})$}\right]f(y,z),
	\label{epsilon=3}
	\ee
where ${\rm erf}$ stands for the error function, and $f$ is a function such that $|f(y,z)|\leq b<1/\sqrt\pi\approx 0.564$ for some $b\in\R^+$. Eq.~(\ref{epsilon=3}) corresponds to setting $u(\fK,y,z)=a \,e^{-a^2\fK^2/4}f(y,z)$ in (\ref{epsilon=}).

\begin{figure}
        \begin{center}
        \includegraphics[scale=.30]{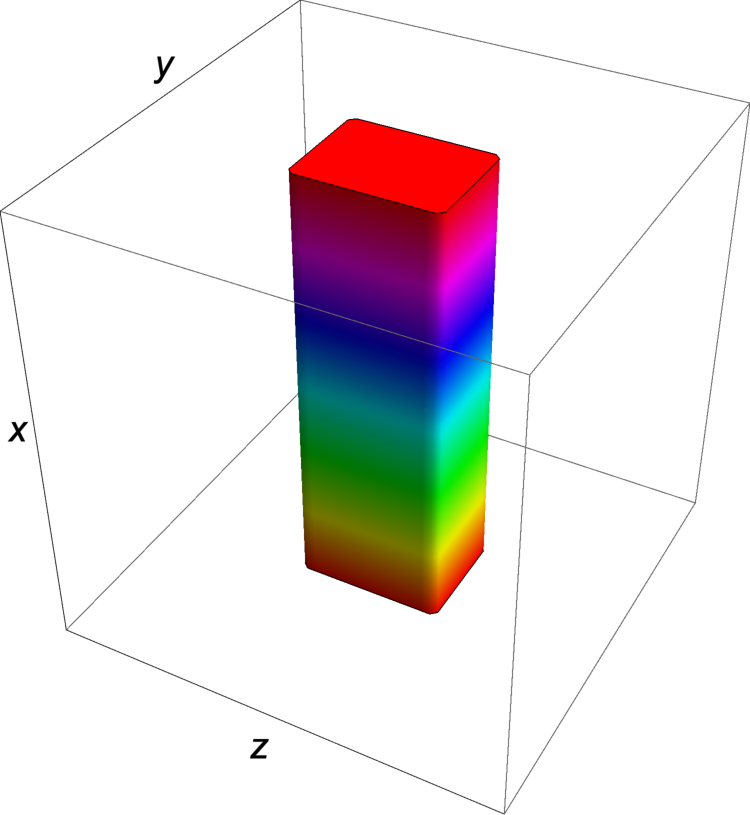}~~~~~~~~~
        \includegraphics[scale=.40]{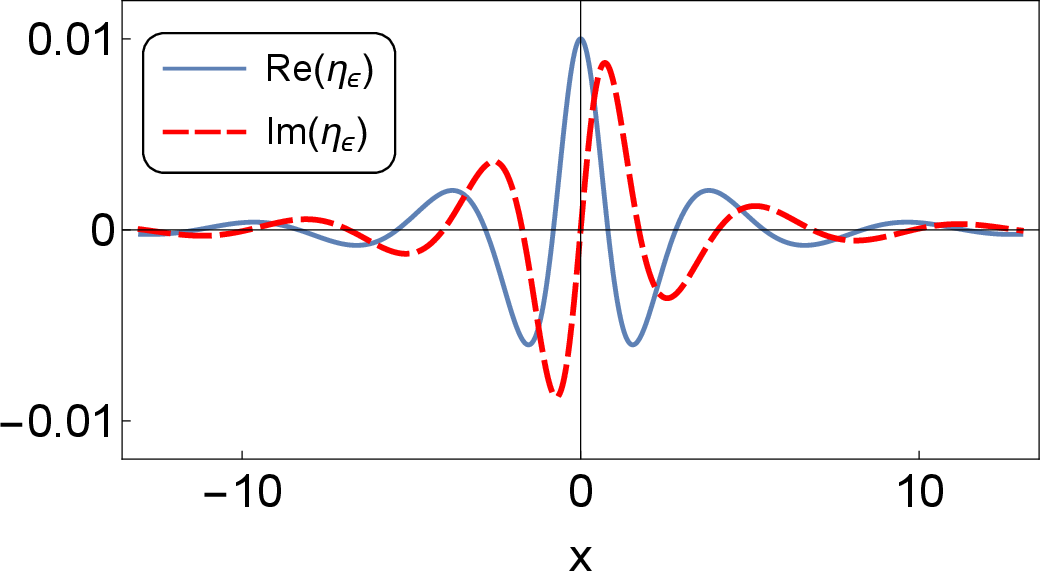}
        \caption{Schematic view of the box confining the inhomogeneous part of the medium given by (\ref{epsilon=2}), (\ref{f=eg}), and (\ref{spec}) (on the left) and the plots of the real and imaginary parts of $\eta_\varepsilon$ as a function of $x$ inside this box (on the right). Here we use units where $\alpha=1$.} 
        \label{fig2}
        \end{center}
        \end{figure}

Consider the following choice of for the function $f$ appearing (\ref{epsilon=2}) and (\ref{epsilon=3}).
	\be
	f(y,z):=\left\{\begin{array}{cc}
	\fz&\for~~|y|\leq \frac{\ell_y}{2}~~{\rm and}~~|z|\leq\frac{\ell_z}{2},\\[6pt]
	0 &{\rm otherwise},\end{array}\right.
	\label{f=eg}
	\ee
where $\fz$ is a nonzero real or complex number, and $\ell_y$ and $\ell_z$ are positive real parameters. Then (\ref{epsilon=2}) and (\ref{epsilon=3}) corresponds to situations where the inhomogeneity of the medium that is responsible for the scattering of waves is confined to an infinite box with a finite rectangular base of side lengths $\ell_y$ and $\ell_z$. The amplitude of the inhomogeneity decays to zero as $|x|\to\infty$ and we can approximate the box by the finite box given by $|x|\leq\frac{\ell_x}{2}$, $|y|\leq \frac{\ell_y}{2}$, and $|z|\leq\frac{\ell_z}{2}$, where $\ell_x$ is a positive real parameter much larger than $a$. Fig.~\ref{fig2} provides a schematic demonstration of this box and plots of real and imaginary parts of $\eta_\varepsilon$ inside the box for the permittivity profile (\ref{epsilon=2}) with 
	\begin{align}
	&\fz=0.01, &&m=1, &&a=2\alpha^{-1}, &&\ell_y=3\alpha^{-1},
	&&\ell_z=4\alpha^{-1}.
	\label{spec}
	\end{align}
Using these numerical values, we find that, $|\eta_\varepsilon(x,y,z)|<10^{-4}$ for $|x|\geq \ell_x=10 a$. Notice also that the broadband invisibility of the permittivity profile given by (\ref{epsilon=2}) and (\ref{spec}) for $k\leq\alpha/2$ remains intact for all real and complex values of $\fz$ such that $|\fz|<1$.
        
We close this section by drawing attention to the following points.
	\begin{itemize}
	\item[-] The hypothesis of Theorem~2 does not prohibit the presence of dispersion, i.e., it also applies to situations where the relative permittivity and permeability tensors depend on the wavenumber $k$. If Conditions (\ref{condi-0}), (\ref{bound}), and (\ref{condi}) hold for all $k\leq\alpha$, the first Born approximation is exact for $k\leq\alpha$, and the medium is invisible for $k\leq \alpha/2$.\footnote{This in turn implies that the medium does not scatter incident wave packets that are superpositions of plane waves with $k\leq\alpha/2$.} For example, the nonmagnetic isotropic media described by (\ref{epsilon=2}) and (\ref{epsilon=3}) satisfy these conditions even if the function $f$ has an arbitrary $k$-dependence. 
	\item[-] We can apply Theorem~2 also for situations where, similarly to the one-dimensional setups considered in Refs.~\cite{horsley-2015,longhi-2015,horsley-longhi},
the regions $z\leq a_-$ and $z\geq a_+$ are filled with a homogeneous and isotropic background medium. In this case we only need to define the incident wavenumber and the relative permittivity and permeability tensors relative to the background medium, i.e., set
$k:=\omega\sqrt{\varepsilon_{\rm B}\mu_{\rm B}}$, $\hat\bfvarepsilon:=\varepsilon_{\rm B}^{-1}\bfvarepsilon$, and $\hat\bfmu:=\mu_{\rm B}^{-1}\bfmu$, where $\varepsilon_{\rm B}$ and $\mu_{\rm B}$ are respectively the permittivity and permeability of the background.
	\end{itemize}

\section{Dynamical formulation of electromagnetic scattering}

Consider a time-harmonic electromagnetic wave propagating in a stationary linear medium with relative permittivity and permeability tensors $\hat\bfvarepsilon$ and $\hat\bfmu$. Then we can express the electric and magnetic fields of the wave in the form $\bsfE(\bfr,t)=e^{-i\omega t}\bcE(\bfr)/\sqrt\varepsilon_0$ and $\bsfH(\bfr,t)=e^{-i\omega t}\bcH(\bfr)/\sqrt\mu_0$, where $\omega$ is the angular frequency of the wave, and $\bcE$ and $\bcH$ are vector-valued functions in terms of which Maxwell's equations take the form \cite{pra-2023}:
	\begin{align}
	&\bnabla\times\bcE=ik\hat\bmu\,\bcH,
	&&\bnabla\times\bcH=-ik\hat\bvarepsilon\,\bcE.
	\label{mx2}
	\end{align}

Suppose that $\hat\varepsilon_{33}\neq 0$ and $\hat\mu_{33}\neq 0$. We can then use (\ref{mx2}) to express the $z$ component of $\bcE$ and $\bcH$ in terms of its $x$ and $y$ components, i.e., $\cE_x,\cE_y,\cH_x$, and $\cH_y$. This in turn allows for reducing (\ref{mx2}) to a system of first-order equations which we can express in the form of the  time-dependent Schr\"odinger equation \cite{pra-2023},
	\be
	i\partial_z\bPhi =\widehat\bH\,\bPhi,
	\label{TD-sch-eq}
	\ee
where $z$ plays the role of time, $\bPhi$ is a 4-component function given by
	\begin{align}
	&\bPhi :=
	\left[\begin{array}{c}
	\vec\cE \\ \vec\cH \end{array}\right],
	&&\vec\cE :=\left[\begin{array}{c}
	\cE_x \\ \cE_y \end{array}\right],
	&&\vec\cH :=\left[\begin{array}{c}
	\cH_x \\ \cH_y \end{array}\right],
	\label{4-com-field}
	\end{align}
and $\widehat\bH$ is a $4\times 4$ matrix-valued differential operator. The latter has the form, 
	\be
	\widehat\bH:=\left[\begin{array}{cc}
	\widehat\bH_{11} & \widehat\bH_{12} \\[6pt]
	\widehat\bH_{21}  & \widehat\bH_{22}\end{array}\right],
	\label{H-def}
	\ee
where $\widehat\bH_{ij}$ are the $2\times 2$ matrix-valued operators given by
	\begin{align}
	&{\widehat\bH}_{11}:=-i{\vec\partial}\,\frac{1}{\hat\varepsilon_{33}}\:\vec\bigvarepsilon_3^{\,T}+
	\frac{1}{\hat\mu_{33}}\:\vec K_\cH\vec\partial^{\,T}\bsigma_2,
    	&&{\widehat\bH}_{12}:=-\frac{i}{k}{\vec\partial}\:\frac{1}{\hat\varepsilon_{33}}
	\vec\partial^{\,T}\bsigma_2 +k\,\bK_\cH,
	\label{H1}\\
    	&{\widehat\bH}_{21}:=\frac{i}{k}{\vec\partial}\:\frac{1}{\hat\mu_{33}}\:
	\vec\partial^{\,T}\bsigma_2- k\, \bK_\cE,
    	&&{\widehat\bH}_{22}:=-i{\vec\partial}\:\frac{1}{\hat\mu_{33}}\:\vec\bigmu_3^{\,T}+
	\frac{1}{\hat\varepsilon_{33}}\vec K_\cE\vec\partial^{\,T}\bsigma_2,
	\label{H2}
	\end{align}
\vspace{-24pt}
	\begin{align}
	&\quad~\vec\partial:=\left[\begin{array}{c}
	\partial_x \\
	\partial_y \end{array}\right],
	&&\bsigma_2:=\left[\begin{array}{cc}
	0 & -i\\
	i & 0 \end{array}\right],
	&&\vec\bigvarepsilon_\ell :=\left[\begin{array}{c}
	\hat\varepsilon_{\ell 1} \\
	\hat\varepsilon_{\ell 2} \end{array}\right],
	&&
	\vec\bigmu_\ell :=\left[\begin{array}{c}
	\hat\mu_{\ell 1} \\
	\hat\mu_{\ell 2} \end{array}\right],
	\label{new}
	\end{align}
$\ell\in\{1,2,3\}$, a superscript ``$T$'' stands for the transpose of a matrix or a matrix-valued operator,  ${\vec\partial}$ and $\vec\partial^{\,T}$ act on all the terms appearing to their right\footnote{For example, for every test function $\bff$,  ${\vec\partial}\,\frac{1}{\hat\varepsilon_{33}}\,\vec\bigvarepsilon_3^{\,T}\bff$ stands for ${\vec\partial} \big(\frac{1}{\hat\varepsilon_{33}}\,\vec\bigvarepsilon_3^{\,T}\bff \big)$.},
and	
	\begin{align}
	&\vec K_\cE:=\Bigg[\begin{array}{c}
	-\hat\varepsilon_{23}\\
	\hat\varepsilon_{13}\end{array}\Bigg],
	&&\bK_\cE:=
	\Bigg[\begin{array}{c}
	-\vec\bigvarepsilon_2^{\,T}\\
	\vec\bigvarepsilon_1^{\,T}\end{array}\Bigg]
	-\frac{1}{\hat\varepsilon_{33}}\vec K_\cE\vec\bigvarepsilon_3^{\,T},
	\label{JEs=}\\
	&\vec K_\cH:=\Bigg[\begin{array}{c}
	-\hat\mu_{23}\\
	\hat\mu_{13}\end{array}\Bigg],
	&&\bK_\cH:=\Bigg[\begin{array}{c}
	-\vec\bigmu_2^{\,T}\\
	\vec\bigmu_1^{\,T}\end{array}\Bigg]
	-\frac{1}{\hat\mu_{33}}\vec K_\cH\vec\bigmu_3^{\,T}.
	\label{JHs=}
	\end{align}
	
The time-dependent Schr\"odinger equation (\ref{TD-sch-eq}) determines the dynamics of an effective quantum system.  Because $z$ plays the role of time, we view $x$ and $y$ as the configuration- (position-) space variables, and identify $\bPhi$ and $\widehat\bH$ respectively with the position wave function for an evolving state and the position-representation of a time-dependent Hamiltonian operator.\footnote{Viewed as an operator acting in the space of 4-component wave functions equipped with the $L^2$-inner product,  $\widehat\bH$ is generally non-Hermitian. This makes the corresponding effective quantum system nonunity.} To make the $z$-dependence of the latter explicit, we denote it by $\widehat H(z)$. Employing Dirac's bracket notation, we can express the evolving state vector by $|\Phi(z)\kt$. By definition, this solves the Schr\"odinger equation,
	\be
	i\partial_z|\Phi(z)\kt=\widehat H(z)|\Phi(z)\kt.
	\ee
We also have $\bPhi(x,y,z)=\bPhi(\vec r,z)=\br\vec r\,|\Phi(z)\kt$ and $\widehat\bH\bPhi(\vec r,z)=\br\vec r\,|\widehat H(z)|\Phi(z)\kt$. We can obtain the explicit form of the Hamiltonian operator $\widehat H(z)$ by making the following changes in the expression for $\widehat\bH$:
$x\to\widehat x$, $y\to\widehat y$, $\partial_x\to i\widehat p_x$, and $\partial_y\to i\widehat p_y$,
where $\widehat x$ and $\widehat y$ are the $x$ and $y$ components of the standard position operators, $\widehat p_x$ and $\widehat p_y$ are the $x$ and $y$ components of the standard momentum operators, and we use conventions where $\hbar=1$.

Let us consider the description of the above effective quantum system in the momentum representation. Because of our convention for the definition of the two-dimensional Fourier transform, i.e., Eq.~(\ref{Fourier}), the momentum wave function associated with the state vector $|\Phi(z)\kt$ is given by 
$\br\vec p\,|\Phi(z)\kt=(2\pi)^{-1}\tilde\bPhi(\vec p,z)$. Denoting the two-dimensional Fourier transform and its inverse respectively by $\cF$ and $\cF^{-1}$, we have $\tilde\bPhi=\cF\bPhi$ and $\bPhi=\cF^{-1}\tilde\bPhi$. The momentum representation of the Hamiltonian, which we label by $\widehat{\tilde\bH}(z)$, satisfies $\widehat{\tilde\bH}(z)\tilde\bPhi(\vec p,z)=2\pi\,\br\vec p\,|\widehat H(z)
	|\Phi(z)\kt$. This in turn shows that $\widehat{\tilde\bH}(z)=\cF\,\widehat{\bH}\,\cF^{-1}$. We can obtain the explicit form of $\widehat{\tilde\bH}(z)$ by making the following changes in the formula for $\widehat{\bH}$: $x\to i\partial_{p_x}$, $y\to i\partial_{p_y}$, $\partial_x\to i p_x$, $\partial_y\to i p_y$.
	
If the wave propagates in vacuum, $\hat\bfvarepsilon=\hat\bfmu=\bI$ and  $\widehat{\tilde\bH}(z)=\widehat{\tilde\bH}_0$, where $\widehat{\tilde\bH}_0\bff(\vec p):=\tilde\bH_0(\vec p)\bff(\vec p)$ and 
	\begin{align}
	&\tilde{\bH}_0(\vec p):=\left[\begin{array}{cc}
    	\bzero & \tilde{\bL}_0(\vec p)\\
    	- \tilde{\bL}_0(\vec p) & \bzero\end{array}\right],
    	&&\tilde{\bL}_0(\vec p):=\frac{1}{k}\left[\begin{array}{cc}
    	-p_xp_y & p_x^2-k^2\\
   	-p_y^2+k^2 & p_xp_y\end{array}\right].
    	\label{t-H-zero}
    	\end{align}  
Because $\widehat{\tilde\bH}_0$ is $z$-independent, its evolution operator has the form, $\widehat{\tilde\bU}_0(z,z_0)=e^{-i(z-z_0)\widehat{\tilde\bH}_0}$, where $z,z_0\in\R$ and $z_0$ represents an initial ``time''. The dynamics generated by $\widehat{\tilde\bH}_0$ corresponds to the propagation of the wave in the absence of the interaction with the medium, i.e., $\widehat{\tilde\bH}_0$ plays the role of a free Hamiltonian in the momentum representation. This suggests that the information about the scattering effects of the medium should be contained in the corresponding interaction-picture Hamiltonian \cite{Sakurai}. In the momentum-space representation, this has the form
	\begin{align}
	\widehat\bsH(z):=e^{iz\widehat{\tilde\bH}_0}\,
	\delta\widehat{\tilde\bH}(z)\,e^{-iz\widehat{\tilde\bH}_0},
	\label{sH}
	\end{align}
where
	\be
	\delta\widehat{\tilde\bH}(z):=\widehat{\tilde\bH}(z)-\widehat{\tilde\bH}_0.
	\label{delta-H}
	\ee	
	
Let $\C^{m\times n}$ denote the space of $m\times n$ complex matrices, $\sF^4$ be the space of $4$-component functions of $\vec p$, and $\sF^4_k$ be the subspace of $\sF^4$ consisting of functions $\bff:\R^2\to\C^{4\times 1}$ such that $\bff(\vec p)=\bzero$ for $|\vec p|<k$. In Ref.~\cite{pra-2023}, we introduce the fundamental transfer matrix $\widehat\bM$ as a linear operator acting in $\sF^4$ that is given by 
	\be
	\widehat\bM= \Lpi\,\widehat\bsU(\infty,-\infty) \Lpi,
	\label{M=}
	\ee
where $\Lpi$ is the projection operator mapping $\sF^4$ onto $\sF^4_k$ according to
	\be
	\big(\Lpi\bff\big)(\vec p):=\left\{\begin{array}{ccc}
	\bff(\vec p)&\for&|\vec p|<k,\\
	\bzero&\for&|\vec p|\geq k,\end{array}\right.
	\label{Lpi-def}
	\ee
and $\widehat\bsU(z,z_0)$ is the interaction-picture evolution operator in the momentum representation.\footnote{Note that $\widehat\bsU(\infty,-\infty)$ coincides with the $S$ matrix of the effective quantum system \cite{weinberg}. See also \cite{pra-2014a}.} Clearly, $\widehat\bM$ maps $\sF^4_k$ to $\sF^4_k$. We can use the Dyson series expansion \cite{Sakurai} of $\widehat\bsU(z,z_0)$ and Eq.~(\ref{M=}) to express it in the form
	\be
	\widehat\bM=\Lpi +\Lpi \sum_{\ell=1}^\infty (-i)^\ell
        \int_{-\infty}^\infty \!\!dz_\ell\int_{-\infty}^{z_\ell} \!\!dz_{\ell-1}
        \cdots\int_{-\infty}^{z_2} \!\!dz_1\,
        \widehat{\bsH}(z_\ell)\widehat{\bsH}(z_{\ell-1})\cdots\widehat{\bsH}(z_1) \Lpi.
        \label{dyson}
        \ee
Notice that if $\bff\in\sF^4_k$, 
	\be
	(\widehat\bM-\bI)\bff=(\widehat\bM-\Lpi)\bff,
	\label{id-202}
	\ee 
where $\widehat\bI$ is the identity operator for $\sF_k^4$.

In order to reveal the relationship between the fundamental transfer matrix and electromagnetic scattering, we make the following observations.

\begin{enumerate}
\item In the coordinate system we have chosen, the source of the incident wave and the detectors are placed on the planes $z=\pm\infty$. The detectors reside on both of these planes, while the source lies on one of them. If the source is on the plane $z=-\infty$ (respectively $z=+\infty$), we say that the wave is left-incident (respectively right-incident). We can quantify these using the spherical coordinates of the incident wave vector $\bk_{\rm i}$ which we denote by $(k,\vartheta_0,\varphi_0)$. For a left-incident wave  $\vartheta_0\in (-\frac{\pi}{2},\frac{\pi}{2})$ and $\cos\vartheta_0>0$. For a right-incident wave $\vartheta_0\in (\frac{\pi}{2},\frac{3\pi}{2})$ and $\cos\vartheta_0<0$. Similarly if we use $(r,\vartheta,\varphi)$ for the spherical coordinate of the position $\bfr$ of a detector placed at $z=+\infty$ (respectively $z=-\infty$) we have $\cos\vartheta>0$ (respectively $\cos\vartheta<0$).

\item Let us introduce
	\begin{align}
	& \bh_{\rm i}:=\frac{1}{k}\,\bk_{\rm i}\times\bfe_{\rm i},
	\quad\quad
	\vec\bfe_{\rm i}:=\left[\begin{array}{c}
	e_{{\rm i}\,x}\\
	e_{{\rm i}\,y}
	\end{array}\right],
	\quad\quad
	\vec\bh_{\rm i}:=\left[\begin{array}{c}
	h_{{\rm i}\,x} \\
	h_{{\rm i}\,y} 
	\end{array}\right],\quad\quad
	\bup_{\rm i}:=\left[\begin{array}{c}
	\vec\bfe_{\rm i}\\
	\vec\bh_{\rm i}\end{array}\right],
	\label{incident}\\[6pt]
	&\vec k_{\rm i}:=(k_{{\rm i}\,x}\,,\,k_{{\rm i}\,y}),
	\quad\quad
	\varpi(\vec p):=\sqrt{k^2-|\vec p|^2},
	\quad\quad
	\sD_k:=\{\vec p\in\R^2\,|\,|\vec p|<k\,\},\\[6pt]
	&\bPi_j(\vec p):=\frac{1}{2}\left[\bI
    	+\frac{(-1)^j}{\varpi(\vec p)}\,\tilde{\bH}_0(\vec p)\right]
	=\frac{1}{2\varpi(\vec p)}\left[
	\begin{array}{cc}
	\varpi(\vec p)&(-1)^j\tilde\bL_0(\vec p)\\
	(-1)^{j+1}\tilde\bL_0(\vec p)&\varpi(\vec p)
	\end{array}\right],
    	\label{proj}
	\end{align}
where subscripts $x$ and $y$ mark the $x$ and $y$ component of the corresponding vector, and $j\in\{1,2\}$ and $|\vec p|\neq k$ in (\ref{proj}). Then it turns out that \cite{pra-2023}
	\begin{align}
	&\bPi_1(\vec k_{\rm i})\bup_{\rm i}=\bup_{\rm i}~~\&~~\bPi_2(\vec k_{\rm i})\bup_{\rm i}=\bzero~~~\for~~~\cos\vartheta_0>0~~(\mbox{left-incident waves}),\label{Pi-Gamma-left}
	\\
	&\bPi_1(\vec k_{\rm i})\bup_{\rm i}=\bzero~~\&~~\bPi_2(\vec k_{\rm i})\bup_{\rm i}=\bup_{\rm i}~~~\for~~~\cos\vartheta_0<0~~(\mbox{right-incident waves}).\label{Pi-Gamma-right}
	\end{align}
It is easy to check that for all $\vec p\in\sD_k$ and $\bGamma\in\C^{4\times 1}$, $\bPi_j(\vec p)\bGamma$ is either zero or an eigenvector of $\bH_0(\vec p)$ with eigenvalue $(-1)^j\varpi(\vec p)$. In view of (\ref{Pi-Gamma-left}) and  (\ref{Pi-Gamma-right}), $\bup_{\rm i}$ is an eigenvector of $\bH_0(\vec k_{\rm i})$ with eigenvalue $-\varpi(\vec k_{\rm i})$ for a left-incident wave (respectively $\varpi(\vec k_{\rm i})$ for a right-incident wave). We can also define a pair of linear projection operators $\widehat\bPi_j$ acting in $\sF^4_k$ according to
	\be
	(\widehat\bPi_j\bff)(\vec p):=\bPi_j(\vec p)\bff(\vec p),
	\ee
where $\vec p\in\sD_k$ and $\bff\in\sF_k^4$. These form an orthogonal pair of projection operators, because $\widehat\bPi_i\widehat\bPi_j=\delta_{ij}\widehat\bPi_j$.

\item In Ref.~\cite{jpa-2020a} we show that the vector-valued function $\bF$ that enters the expression (\ref{Scattered-wave-lr}) for the electric field of the scattered wave is given by 
	\be
	\bF(\bk_{\rm s},\bk_{\rm i})=-\frac{ik|\cos\vartheta|}{2\pi}\:
   	\bXi^T\bT_\pm(\vec k_{\rm s}\,)
	~~~~~~{\rm for}~~~\pm\cos\vartheta>0,
	\label{soln}
	\ee
where $\bXi^T$ is a $1\times 4$ matrix with entries belonging to $\R^3$ that is given by
	\be
	\bXi^T:=\left[\begin{array}{cccc}
	\bfe_x~ &
	~\bfe_y~ &~
	\sin\theta\sin\varphi\,\bfe_z~ &
	~-\sin\theta\cos\varphi\,\bfe_z\end{array}\right],
	\label{bXi-def}
	\ee
$\bfe_x,\bfe_y$, and $\bfe_z$ are respectively unit vectors along the $x,y$, and $z$ axes, $\bT_\pm\in\sF^4_k$ are the 4-component functions satisfying
	 \begin{align}
    	&\widehat\bPi_1\bT_+=\bT_+,
    	\quad\quad\widehat\bPi_2\bT_-=\bT_-,
     	\quad\quad\widehat\bPi_1\bT_-=\widehat\bPi_2\bT_+=\bzero,
    	\label{id-21}\\[6pt]
	&\widehat\bPi_2\,\widehat\bM\,\bT_-=
	-4\pi^2\widehat\bPi_2\big(\widehat\bM-\bI\big)
	\bup_{\rm i} \delta_{\vec k_{\rm i}},
	\label{Eq-TmL}\\[6pt]
	&\bT_+=\widehat\bPi_1\big(\,\widehat{\bM}-\widehat\bI\big)\big(\bT_-+4\pi^2
	\bup_{\rm i} \delta_{\vec k_{\rm i}}\big),
	\label{Eq-TpL}
	\end{align}
$\delta_{\vec k_{\rm i}}$ is the Dirac delta function in two dimensions centered at $\vec k_{\rm i}$, i.e., $\delta_{\vec k_{\rm i}}(\vec p):=\delta(\vec p-{\vec k_{\rm i}})$, and $\vec k_{\rm s}$ is the projection of $\bk_{\rm s}$ onto the $x$-$y$ plane. Note that
	\begin{align}
	&\vec k_{\rm i}:=k\sin\vartheta_0(\cos\varphi_0\,\bfe_x+\sin\varphi_0\,\bfe_y),
	&&\vec k_{\rm s}:=k\sin\vartheta(\cos\varphi\,\bfe_x+\sin\varphi\,\bfe_y).\nn
	\end{align}
	
\end{enumerate}
	
Equation (\ref{Eq-TpL}) specifies $\bT_+$ in terms of $\widehat\bM$ and $\bT_-$. Equation (\ref{Eq-TmL}) is a linear integral equation for $\bT_-$. Dynamical formulation of stationary electromagnetic scattering reduces the scattering problem (finding $\bF$) to the calculation of the fundamental transfer matrix and the solution of (\ref{Eq-TmL}). Substituting the solution of this equation in (\ref{Eq-TpL}) and using (\ref{Scattered-wave-lr}) and (\ref{soln}), we obtain the electric field of the scattered wave. Refs.~\cite{pra-2023,jpa-2020a} offer concrete applications of this approach in the study of electromagnetic point scatterers and the construction of isotropic scatterers that display broadband omnidirectional invisibility.

\section{Exactness of the first Born approximation and broadband invisibility}

Theorem~1 provides a necessary and sufficient condition for the exactness of the first Born approximation in electromagnetic scattering. We will prove Theorem~2 by showing that its hypothesis implies this condition. This requires some preparation.
We begin by introducing some useful notation: 
	\begin{itemize}
	\item[-] Given a function $f:\R^3\to\C$, $\eta_f$ stands for $f-1$, i.e., $\eta_f(\bfr):=f(\bfr)-1$. 
	\item[-] For each $d\in\Z^+$, we use $\fF_d$ to denote the space of functions $\phi:\R^d\to\C$, and label the $d$-dimensional Fourier transform of $\phi$ by $\breve\phi$.
	\item[-] For each $\alpha\in\R$, $\cS_\alpha:=\{\,\ff\in\fF_1\,|\,\ff(p)=0~\for~p\leq\alpha\}$.
	\item[-] Given $v\in\fF_3$ and $z\in\R$, we introduce the operator $\widehat\sV(z):=v(\widehat x,\widehat y,z)$ which acts in the space of functions of $\vec p:=(p_x,p_y)$ according to
	\begin{align}
	\big(\widehat\sV(z)\phi\big)(\vec p)&=\br \vec p\,|v(\widehat x,\widehat y,z)|\phi\kt
	=\frac{1}{4\pi^2}\int_{\R^2} d^2q\:\tilde v(\vec p-\vec q,z)\phi(\vec q),
	\label{sV=}
	\end{align}
where $\tilde v:=\cF v$, i.e., $\tilde v(\vec p,z):=\int_{\R^2} d^2r\: e^{-i\vec r\cdot\vec p}v(\vec r,z)$. Note also that because $\br\vec p|\widehat x|\phi\kt=i\partial_{p_x}\phi(\vec p)$ and $\br\vec p|\widehat y|\phi\kt=i\partial_{p_y}\phi(\vec p)$, we have 
	\[\big(\widehat\sV(z)\phi\big)(\vec p)=v(i\partial_{p_x},i\partial_{p_y},z)\phi(\vec p).\]

	\item[-] For each $k\in\R^+$, let $\sLpi:\fF_2\to\fF_2$ be the operator defined by
		\be
		(\sLpi\phi)(\vec p):=\left\{\begin{array}{ccc}
		\phi(\vec p)&\for&|\vec p|<k,\\
		0&\for&|\vec p|\geq k.
		\end{array}\right.
		\label{sLpi-def}
		\ee
	\end{itemize}
The following lemma lists some of the immediate consequences of the definitions of $\cS_\alpha$ and $\sLpi$.
	\begin{itemize}
	\item[]{\bf Lemma 1}: Suppose that $\alpha,\beta\in\R$ such that $\alpha\leq\beta$, $k\in\R^+$, and $\ff,\fg\in\fF_1$. Then
	\begin{enumerate}
	\item $\cS_\beta\subseteq\cS_\alpha$.
	\item $\sLpi\ff\in\cS_{-k}$.
	\item If $k\leq\alpha$ and $\ff\in\cS_\alpha$, $\sLpi\ff=0$.
	\item If $\ff\in\cS_\alpha$, $\ff\,\fg\in\cS_\alpha$. In particular, if $\fg\in\cS_\beta$, $\ff\,\fg\in\cS_\beta$. 
	\end{enumerate} 
	\end{itemize}
The following two lemmas reveal less obvious facts about $\cS_\alpha$. We give their proofs in Appendix~C of Ref.~\cite{jpa-2020a}.
	\begin{itemize}
	\item[]{\bf Lemma 2}: Let $\alpha\in\R$ and $\phi_1,\phi_2,\phi_3\in\fF_1$ be such that 
	$\breve\phi_1,\breve\phi_2\in\cS_\alpha$ and $\phi_3=\phi_1\phi_2$. Then $\breve\phi_3\in\cS_{2\alpha}\subseteq \cS_{\alpha}$.
	\end{itemize}
	\begin{itemize}
	\item[]{\bf Lemma 3}: Let $\alpha\in\R$ and $f:\R\to\C$ be a bounded function whose real part is bounded below by a positive number, i.e., there are $m,M\in\R$ such for all $x\in\R$, $0<m\leq\RE[f(x)]\leq|f(x)|\leq M$. Then there is a sequence of complex numbers $\{c_n\}_1^\infty$ such that the series $\sum_{n=1}^\infty c_n\eta_{_f}(x)^n$ converges absolutely to $\eta_{_{1/f}}(x)$, so that $\eta_{_{1/f}}=\sum_{n=1}^\infty c_n\eta_{_f}^n$. Furthermore if $\eta_{_f},\eta_{_{1/f}}\in\fF_1$ and $\breve\eta_{_f}\in\cS_\alpha$, we have $\breve\eta_{_{1/f}}\in\cS_\alpha$.
	\end{itemize}
We can use Lemmas 2 and 3 to establish:	
	\begin{itemize}
	\item[]{\bf Lemma 4}: Let $f$ be as in Lemma~3, $g\in\fF_1$, $h:=g/f$, and $\alpha\in\R$. Suppose that $\breve\eta_{_f}\in\cS_\alpha$ and $\breve g\in\cS_\alpha$. Then $\breve h \in\cS_\alpha$.
	\item[] Proof: Lemma~2 implies $h=g+\sum_{n=1}^\infty c_n\eta_f^ng$. This equation together with Lemma~2 and the conditions $\breve\eta_{_f}\in\cS_\alpha$ and $\breve g\in\cS_\alpha$ imply $\breve h \in\cS_\alpha$.
	\end{itemize}
In Appendix~B of Ref.~\cite{pra-2021}, we prove the following lemma.
	\begin{itemize}
	\item[]{\bf Lemma 5}: Let $\phi\in\fF_2$, $v\in\fF_3$, $z\in\R$, $\widehat\sV(z):=v(\widehat x,\widehat y,z)$, $\psi\in\fF_2$, and $\alpha,\beta\in\R$. Suppose that for all $p_y,z\in\R$, $\psi(\cdot,p_y)\in\cS_\alpha$ and $\tilde v(\cdot,p_y,z)\in\cS_\beta$. Then $\widehat\sV(z)\psi\in\cS_{\alpha+\beta}$.		
	\end{itemize}
Next, we present a variation of Lemma~4 of Appendix~B of Ref.~\cite{pra-2021} which follows from the same argument.
	\begin{itemize}
	\item[]{\bf Lemma 6}: Let $\alpha\in\R^+$, $\beta\in\R$, $k\in (0,\alpha]$, $\phi\in\fF_2$, $n\in\Z^+$, for all $i\in\{1,2,\cdots,n\}$,  $z_i\in\R$, $v_i\in\fF_3$, and $\widehat\sV_i(z_i):=v_i(\widehat x,\widehat y,z_i)$, for all $j\in\{0,1,2,\cdots,n\}$, $\xi_j\in\fF_2$ and $\widehat\xi_j:=\xi_j(\widehat p_x,\widehat p_y)$, and 
		\[\phi_n:=\widehat\xi_n\widehat\sV_n(z_n)\,\widehat\xi_{n-1}\widehat\sV_{n-1}(z_{n-1})
		\,\widehat\xi_{n-2}\cdots \widehat\xi_1\widehat\sV_1(z_1)\,
		\widehat\xi_0\,\sLpi\phi.\]
Suppose that for all $p_y,z\in\R$ and $i\in\{1,2,\cdots,n\}$, $\tilde v_i(\cdot,p_y,z)\in\cS_\beta$. Then $\phi_n(\cdot,p_y)\in\cS_{n\beta-\alpha}$. In particular,  
	\be
	\sLpi\,\widehat\xi_n\widehat\sV_n(z_n)\,\widehat\xi_{n-1}\widehat\sV_{n-1}(z_{n-1})
		\,\widehat\xi_{n-2}\cdots \widehat\xi_1\widehat\sV_1(z_1)\,
		\widehat\xi_0\,\sLpi
	\label{Lemma6-e1}
	\ee
coincides with the zero operator $\hat 0$ if $\beta\geq\frac{2\alpha}{n}$.
	\end{itemize}
According to this lemma, setting $\beta=2\alpha$ and $\beta=\alpha$ we respectively obtain 
	\begin{align}
	&\sLpi\,\widehat\xi_1\widehat\sV_1(z_1)\,
		\widehat\xi_0\,\sLpi=\widehat 0~~~~{\rm if}~~~~\tilde v_1(p_x,p_y,z)=0~~\for~~ p_x\leq 2\alpha,
	\label{condi-11}\\
	&\sLpi\,\widehat\xi_2\widehat\sV_2(z_2)\,\widehat\xi_1\widehat\sV_1(z_1)\,
		\widehat\xi_0\,\sLpi=\widehat 0~~~{\rm if}~~~
		\tilde v_1(p_x,p_y,z)=\tilde v_2(p_x,p_y,z)=0~~\for~~ p_x\leq \alpha.
	\label{condi-12}
	\end{align}
 		
Next, we examine the operator $\delta\widehat{\tilde\bH}(z)$ of Eq.~(\ref{delta-H}). Clearly, $\delta\widehat{\tilde\bH}(z)=\cF\, \delta\widehat\bH(z)\cF^{-1}$, where $\delta\widehat\bH(z):=\widehat\bH(z)-\widehat\bH_0$. The fact that $\widehat\bH_0$ is obtained from $\widehat\bH(z)$ by setting $\hat\bfvarepsilon=\hat\bfmu=\bI$ together with Eqs.~(\ref{H-def}) -- (\ref{JHs=}) show that we can obtain $\delta\widehat\bH(z)$ from the expression for $\widehat\bH(z)$ by making the following changes.
	\begin{itemize}
	\item[-] In Eqs.~(\ref{H1}) and (\ref{H2}): $\frac{1}{\hat\varepsilon_{33}}\to\eta_{\,{1/\hat\varepsilon_{33}}}$ and $\frac{1}{\hat\mu_{33}}\to\eta_{\, {1/\hat\mu_{33}}}$;
	\item[-] In Eqs.~(\ref{new}) -- (\ref{JHs=}): $\hat\varepsilon_{ij}\to\eta_{\bfvarepsilon,ij}$ and $\hat\mu_{ij}\to\eta_{\bfmu,ij}$, where $\eta_{\bfvarepsilon,ij}:=\hat\varepsilon_{ij}-\delta_{ij}$, $\eta_{\bfmu,ij}:=\hat\mu_{ij}-\delta_{ij}$, and $\delta_{ij}$ stands for the Kronecker delta symbol.
	\end{itemize}
Employing this prescription to determine the entries of $\delta\widehat\bH(z)$ and making use of Lemma 6, we find that whenever Condition~(\ref{condi2}) holds, the Fourier transform with respect to $x$ of all the functions appearing in the expression for $\delta\widehat\bH(z)$ vanish for $p_x\leq \alpha$. Furthermore, we can use (\ref{sH}),
to infer that the entries of $\widehat{\bsH}(z_2)\widehat{\bsH}(z_1)\Lpi$ are sums of the terms of the form (\ref{Lemma6-e1}). This together with (\ref{Lpi-def}), (\ref{sLpi-def}), (\ref{condi-12}), and the fact that $e^{\pm i z\widehat{\tilde\bH}_0}$ and $\Lpi$ commute imply that the quadratic and higher order terms of the Dyson series (\ref{dyson}) vanish. Therefore,
	\be
	\widehat\bM=\Lpi-i \int_{-\infty}^\infty \!\!dz\; \Lpi\widehat{\bsH}(z_1)\Lpi=
	\Lpi-i \int_{-\infty}^\infty \!\!dz\;  e^{i z\widehat{\tilde\bH}_0}\Lpi
	\delta\widehat{\tilde\bH}(z)\Lpi e^{- i z\widehat{\tilde\bH}_0}.
        \label{M-2}
        \ee
Substituting the explicit form of $\delta\widehat{\tilde\bH}(z)$ in $\Lpi\,\delta\widehat{\tilde\bH}(z)\Lpi$, we find that its entries are sums of terms of the form (\ref{Lemma6-e1}) which vanish unless they involve one and only one of $\eta_{\bfvarepsilon,ij}$ and $\eta_{\bfmu,ij}$. This implies that
	\be
	\Lpi\,\delta\widehat{\tilde\bH}(z)\Lpi=\Lpi\left[\begin{array}{cc}
	\widehat\bV_{11} & \widehat\bV_{12}\\
	\widehat\bV_{21} & \widehat\bV_{22}\end{array}\right]\Lpi,
	\label{Lpi-H=Lpi}
	\ee
where
	\begin{align}
	&\hspace{-12pt}\widehat\bV_{11}:={\vec P}\: \delta\widehat{\vec\bigvarepsilon}_3^T+
	i\,\delta\widehat{\vec K}_\cH {\vec P}^T\!\!\bsigma_2,
	&& \widehat\bV_{12}:=-\frac{i}{k}{\vec P}\: \widehat\eta_{\bfvarepsilon_{33}}
	{\vec P}^T\!\!\bsigma_2+k\,\delta\widehat\bK_\cH,
	\label{delta-1}\\
	&\hspace{-12pt}\widehat\bV_{21}:=\frac{i}{k}{\vec P}\: \widehat\eta_{\bfmu_{33}}
	{\vec P}^T\!\!\bsigma_2-k\,\delta\widehat\bK_\cE,
	&& \widehat\bV_{22}:={\vec P}\: \delta\widehat{\vec\bigmu}_3^T+
	i\,\delta\widehat{\vec K}_\cE {\vec P}^T\!\!\bsigma_2,
	\label{delta-2}
	\end{align}\vspace{-24pt}
	\begin{align}
	&\quad\quad{\vec P}:=\left[\begin{array}{c}
	p_x\\
	p_y\end{array}\right],
	&&\quad\quad
	\delta\widehat{\vec\bigvarepsilon}_\ell:=
	\left[\begin{array}{c}
	\widehat\eta_{\bfvarepsilon,\ell 1}\\
	\widehat\eta_{\bfvarepsilon,\ell 2}\end{array}\right],
	&&
	\quad\quad
	\delta\widehat{\vec\bigmu}_\ell:=
	\left[\begin{array}{c}
	\widehat\eta_{\bfmu,\ell 1}\\
	\widehat\eta_{\bfmu,\ell 2}\end{array}\right],
	\label{delta-3}\\
	&\quad\quad
	\delta\widehat{\vec K}_\cE:=\left[\begin{array}{c}
	-\widehat\eta_{\bfvarepsilon,23}\\
	\widehat\eta_{\bfvarepsilon,13}\end{array}\right],
	&&\quad\quad
	\delta\widehat{\bK}_\cE=\left[\begin{array}{c}
	-\delta\widehat{\vec\bigvarepsilon}_2^{_{\mbox{\scriptsize$\,T$}}}\\
	\delta\widehat{\vec\bigvarepsilon}_1^{_{\mbox{\scriptsize$\,T$}}}\end{array}\right],
	&&\quad\quad
	\delta\widehat{\vec K}_\cH:=\left[\begin{array}{c}
	-\widehat\eta_{\bfmu,23}\\
	\widehat\eta_{\bfmu,13}\end{array}\right],
	\label{delta-4}\\
	&\quad\quad
	\delta\widehat{\bK}_\cH=\left[\begin{array}{c}
	-\delta\widehat{\vec\bigmu}_2^{_{\mbox{\scriptsize$\,T$}}}\\
	\delta\widehat{\vec\bigmu}_1^{_{\mbox{\scriptsize$\,T$}}}\end{array}\right],
	&& 
	\widehat\eta_{\bfvarepsilon,ij}:=\eta_{\bfvarepsilon,ij}(i\partial_{p_x},i\partial_{p_y},z),
	&& 
	\widehat\eta_{\bfvarepsilon,ij}:=\eta_{\bfmu,ij}(i\partial_{p_x},i\partial_{p_y},z),
	\label{delta-5}
	\end{align}
and we have also benefitted from Lemmas~3 and 4.

In view of the argument leading to (\ref{M-2}), Condition~(\ref{condi2}), and the fact that $\Lpi$ commutes with $\widehat\bPi_2$, we have
	\be
	\big(\widehat\bM-\Lpi\big)\widehat\bPi_2
	\big(\widehat\bM-\Lpi\big)=\widehat\bzero,
	\label{id-101}
	\ee
where $\widehat\bzero$ is the zero operator acting in $\sF^4$. This identity allows us to solve Eq.~(\ref{Eq-TmL}) for $\bT_-$. To see this, we use (\ref{id-202}) and (\ref{id-21}) to write (\ref{Eq-TmL}) in the form
	\be
	\bT_-=-\widehat\bPi_2(\widehat\bM-\Lpi)(\bT_-+4\pi^2
	\bup_{\rm i} \delta_{\vec k_{\rm i}}).
	\label{T-minus-1}
	\ee	
Applying $\widehat\bM-\Lpi$ to both sides of this equation and making use of (\ref{id-101}), we obtain $(\widehat\bM-\Lpi)\bT_-=\bzero$. Substituting this relation in (\ref{Eq-TmL}) and (\ref{T-minus-1}), we are led to
	\bea
	\bT_+&=&4\pi^2\widehat\bPi_1\big(\widehat\bM-\Lpi\big) 
	\bup_{\rm i} \delta_{\vec k_{\rm i}},
	\label{T-plus=}\\
	\bT_-&=&-4\pi^2\widehat\bPi_2\big(\widehat\bM-\Lpi\big) 
	\bup_{\rm i} \delta_{\vec k_{\rm i}}.
	\label{T-minus=}
	\eea

Next, we examine the transformation property of $\bT_\pm$ under (\ref{scaling}). In view of (\ref{M-2}) -- (\ref{delta-5}), (\ref{T-plus=}), and (\ref{T-minus=}),  the scaling transformation (\ref{scaling}) implies $\widehat\bM-\Lpi \to \varsigma\big(\widehat\bM-\Lpi\big)$ and $\bT_\pm \to \varsigma\,\bT_\pm$.
Using this in (\ref{soln}), we find that the electric field of the scattered wave (\ref{Scattered-wave-lr}) scales as $\bE_s\to\varsigma\,\bE_s$. By virtue of Theorem~1, this establishes the exactness of the first Born approximation.

To arrive at a direct proof of the exactness of the first Born approximation, we have substituted (\ref{M-2}) in (\ref{T-plus=}) and (\ref{T-minus=}), and used (\ref{Scattered-wave-lr}), (\ref{soln}), and 
(\ref{delta-1}) -- (\ref{delta-5}) to determine the explicit form of $\bE_s$.  After lengthy calculations we have shown that the resulting formula for $\bE_s$ coincides with the one obtained by performing the first Born approximation, namely the one given by Eqs.~4.18 and 4.29 of Ref.~\cite{newton}. This provides a highly nontrivial check on the validity of our analysis.

For incident waves with wavenumber $k\leq\alpha/2$, we can use (\ref{condi-11}) to show that $\Lpi\delta\widehat{\tilde\bH}(z)\Lpi=\widehat\bzero$. Therefore, $\widehat\bM=\Lpi$, and  (\ref{T-plus=}) and (\ref{T-minus=}) give $\bT_\pm=\bzero$. In view of (\ref{Scattered-wave-lr}) and (\ref{soln}), this implies $\bF=\bE_{\rm s}=\bzero$ which means that the medium does not scatter the wave. Since this result is not sensitive to the direction of the incident wave vector, the medium is omnidirectionally invisible in the wavenumber spectrum $(0,\alpha/2)$. This extends a result of Ref.~\cite{jpa-2020a} to anisotropic media.

\section{Concluding remarks}

The Born approximation has been an indispensable tool for performing quantum and electromagnetic scattering calculations since its introduction in 1926 \cite{Born-1926}. It is therefore rather surprising that the discovery of conditions for its exactness had to wait till 2019 when such a condition was found in the context of the dynamical formulation of stationary scattering for scalar waves in two dimensions \cite{pra-2019}. This condition emerged in an attempt to truncate the Dyson series for the fundamental matrix. It turned out to allow for an exact solution of the scattering problem leading to a formula that was identical to the one obtained by the first Born approximation. The extension of this condition to potential scattering in three dimensions is rather straightforward \cite{pra-2021}. This is by no means true for its generalization to electromagnetic scattering because of the transverse vectorial nature of electromagnetic waves and tensorial nature of the interaction potentials $\bfeta_\bfvarepsilon$ and $\bfeta_\bfmu$. Progress in this direction required the development of a dynamical formulation of stationary electromagnetic scattering which was realized quite recently \cite{pra-2023}. 

The condition for the exactness of the first Born approximation for the scattering of electromagnetic waves shares the basic features of the corresponding condition in potential scattering, and it is quite simple to state and realize. Yet establishing the fact that this condition actually implies the exactness of the first Born approximation requires overcoming serious technical difficulties.

The discovery of a sufficient condition for the exactness of the first Born approximation may be viewed as basic but at the same time formal contribution to the vast subject of scattering theory. One must however note that systems satisfying this condition are exactly solvable. Therefore, imposing this condition yields a very large class of exactly solvable scattering problems.  As it should be clear from the two examples we have provided in Sec.~2, it is possible to satisfy this condition for permittivity and permeability profiles whose expressions involve arbitrary functions of two of the coordinates, e.g., the function $f(y,z)$ of Eqs.~(\ref{epsilon=2}) and (\ref{epsilon=3}). In principle, one can choose these functions so that the system has certain desirable scattering features. Because the formula given by the first Born approximation specifies the scattered wave in terms of the three-dimensional Fourier transform of the relative permittivity and permeability tensors \cite{newton}, one can determine the specific form of $\hat\bfvarepsilon$ and $\hat\bfmu$ by performing inverse Fourier transform of the scattering data. This corresponds to an electromagnetic analog of a well-known approximate inverse scattering scheme for scalar waves that relies on the first Born approximation \cite{Devaney-1982,Chadan-IS}. If one manages to enforce the condition we have provided for the exactness of the first Born approximation, this scheme becomes exact. This suggests that our results may be used to develop a certain exact but conditional inverse scattering scheme. The study of the details and prospects of this scheme is the subject of a future investigation.

\vspace{12pt}
\noindent{\bf Acknowledgements}:
This work has been supported by the Scientific and Technological Research Council of T\"urkiye (T\"UB\.{I}TAK) in the framework of the project 120F061 and by Turkish Academy of Sciences (T\"UBA).

\ed